\documentstyle[epsf,twocolumn]{jpsj}

\def\JH{J_{\rm H}}

\catcode`\@=11
\def\simle{\mathrel{\mathpalette\@versim<}}   % < over \sim
\def\simge{\mathrel{\mathpalette\@versim>}}   % > over \sim
\def\@versim#1#2{\lower2.5pt\vbox{\baselineskip0pt \lineskip-.5pt
   \ialign{$\m@th#1\hfil##\hfil$\crcr#2\crcr\sim\crcr}}}
\catcode`\@=12

\title
{Magnon Broadening Effect by Magnon-Phonon Interaction in 
Colossal Magnetoresistance Manganites}

\author{Nobuo {\sc Furukawa}}

\def\runtitle{Magnon Broadening Effect by Magnon-Phonon Interaction}
\def\runauthor{Nobuo {\sc Furukawa}}

\inst
{
Department of Physics, Aoyama Gakuin University, 
Setagaya, Tokyo 157-8572
}

\recdate
{
\today
}

\abst
{
In order to study the magnetic excitation behaviors
in colossal magnetoresistance manganites,
a magnon-phonon interacting system is investigated.
Sudden broadening of magnon linewidth is obtained
when a magnon branch crosses over an optical phonon branch.
Onset of the broadening is approximately determined by the
magnon density of states.
Anomalous magnon damping at the brillouine zone boundary 
observed in low Curie temperature manganites is explained.
}

\kword
{
Colossal magnetoresistance manganites, zone boundary magnon broadening,
magnon-phonon interaction
}

\begin{document}
\sloppy
\maketitle

{\em Introduction:}
One of the main interests in colossal magnetoresistance (CMR) manganites
is to investigate changes in the microscopic behaviors
upon the bandwidth control.\cite{Furukawa99}
Compounds with larger ionic radius for A-site ions have
relatively high Curie temperature ($T_{\rm c}$) and exhibit
canonical half-metallic double exchange (DE)\cite{Zener51} behaviors.
This  is considered to be due to large $e_{\rm g}$ electron hopping
and thus wide bandwidth of the itinerant electrons which 
favors the ferromagnetic state to maximizes the kinetic energy.
On the other hand, decreasing the radius makes $T_{\rm c}$ lower and
finally drives the system into charge-ordered 
insulator.\cite{Hwang95,Ramirez97}
Magnetoresistance phenomena becomes prominent in the intermediate
region where ferromagnetism and charge ordering are competing.
In order to study the effect of such competitions to various
physical properties, it is important to make an 
investigation  of these compounds in the intermediate region.

One example of such systematic studies
 is the measurement of magnon dispersion relations of these
compounds in the ferromagnetic state.
For high-$T_{\rm c}$ compounds, {\em e.g.} (La,Sr)MnO$_3$ and (La,Pb)MnO$_3$
with hole-carrier
 concentration $x\sim 0.3$, cosine-type dispersion relations are 
reported.\cite{Perring96,Moudden98} 
This behavior is reproduced by the linear spin-wave 
approximation\cite{Kubo72,Furukawa96} of the single-band DE
model at $S\to\infty$ and $\JH\to\infty$,
where $S$ is the $t_{2\rm g}$ spin and $\JH$ is the Hund's coupling,
respectively.
For low-$T_{\rm c}$ compounds
(Pr,Sr)MnO$_3$, (Nd,Sr)MnO$_3$ and (La,Ca)MnO$_3$, however,
magnon dispersion is reported to deviate from the
 cosine-band.\cite{Hwang98,Dai99x}
Prominent softening of magnon at the zone boundary is observed.

Although  zone boundary softening occurs at
finite $\JH$ and finite $S$ 
in the single-band DE model,\cite{Furukawa96,Kaplan97}
it is not sufficiently large compared to the data.
Taking into account the realistic electronic structure,
Solovyev {\em et al.}\cite{Solovyev99} 
discuss that the spin wave softening is of purely magnetic origin.
It is claimed to be essential to include the Mn $t_{2\rm g}$ and O $2p$ 
orbitals.\cite{Mahadevan99x}
On the other hand, Khalliullin {\em et al.}\cite{Khaliullin99x}
have calculated the spin wave dispersion in the presence of orbital
fluctuations as well as phonons to explain the zone boundary softening.
In such a case, softening occurs as a precursor of
ferromagnetic orbital ordering accompanied by the $A$-type antiferromagnetism.

Another point of issue is broadening effect of the magnons.
In wide band compounds (La,Sr)MnO$_3$ and (La,Pb)MnO$_3$ with high $T_{\rm c}$
($x=0.2\sim0.3$),
magnon dispersion is observed throughout the brillouine 
zone at low temperature limit.\cite{Perring96,Moudden98}
There is no magnon damping at high magnon-frequency region
expected for itinerant weak ferromagnets.
Within the DE model with large $\JH$, the
density of states for itinerant $e_{\rm g}$ electrons
shows a large exchange splitting.\cite{Furukawa96}
The Stoner excitation is gapped and its continuum
lies at $\omega \sim 2\JH$, which does not
interact with low energy magnons.
Magnon damping due to thermally induced minority band scales
as $\Gamma \propto (1-m^2)\cdot \omega_q$,
where $m = M_{\rm tot}/ M_{\rm sat}$ is the total magnetization ($M_{\rm tot}$)
scaled by the saturation magnetization at the lowest temperature
($M_{\rm sat}$). 
This scaling form is indeed observed in the wide bandwidth compound
at small $q$.\cite{Furukawa98}

However, in narrow band compound (Pr,Sr)MnO$_3$,
sudden broadening of the magnon linewidth near 
the zone boundary is observed.\cite{Hwang98} 
More quantitative measurement has been done for 
another low $T_{\rm c}$ compound,
(La,Sr)MnO$_3$ with less doping $x=0.15$.\cite{VasiliuDoloc98}
The inelastic neutron scattering data
show narrow magnon linewidth at small $q$ region
and  sudden increase of linewidth near the zone boundary.
The experiment is performed  at low temperature with saturated magnetization 
$m \sim1$ where no Stoner contribution is expected.
Indeed, in the small $q$ region magnon linewidth obeys the scaling law due to
the magnon-magnon scattering,
$\Gamma_{\rm mag} \propto q^4 {\rm ln}^2(T/\omega_q)$.
On the other hand, magnons at the zone boundary
have far wider linewidth than the scaling form.
They are heavily damped so that the linewidth are
 comparable to the magnon energy.
Such magnon damping should be attributed to some mechanisms other than
Heisenberg-type interactions.
In the case of (Nd,Sr)MnO$_3$ at $x=0.3$, zone boundary magnons are
overdamped and below the sensitivity of the measurement.\cite{Dai99x}

Investigation of the mechanisms for such anomalous magnon damping 
is very important to understand the magnetic and electronic
 behaviors of CMR manganites in the low $T_{\rm c}$ intermediate 
bandwidth region.
This may cast doubt on adapting the half-metallic electronic 
structure\cite{Park98a}
for these low $T_{\rm c}$ 
compounds if the damping mechanism is due to Stoner continuum absorption. 
Another possibility is that strong spin-orbital exchange 
interaction\cite{Kugel73} causes the damping of magnon
coupled  to the orbital liquid states.\cite{Ishihara97}

Recent experiment by Dai {\em et al.}\cite{Dai99x} shows that
the onset of the linewidth broadening
 coincides with the point where the
magnon branch crosses the longitudinal optical phonon branch,
for various directions in the brillouine zone.
Magnon linewidth below the crossing frequency is
narrow. 
Therefore, they emphasized the role of magnon-phonon coupling
as a mechanism of the anomalous magnon broadening.
Another important aspect observed here is that 
phonon broadening is not reported
in the entire the brilloiune zone, even though
strong magnon-phonon coupling is speculated.
In this letter, we focus on the role of the magnon-phonon interaction
and calculate the linewidth broadening of magnons and phonons.

\begin{figure}[htb]
\hfil\epsfxsize=6cm\epsfbox{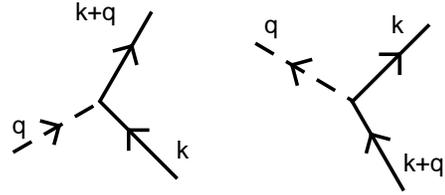}\hfil
\caption{Magnon-phonon vertices. Solid lines and dashed lines
represent magnon and phonon propagators, respectively.}

\label{FigVertex}
\end{figure}

{\em Magnon-Phonon interaction:}
In  CMR manganites,
cooperative Jahn-Teller effect which strongly couples
lattice distortion to conduction electrons
has been discussed.\cite{Kanamori59}
In such case, magnon-phonon coupling will be produced
by the modulation of the DE interaction $J$ by  lattice displacements
in the form
$J(x) = J(x_0) + (\partial J/\partial x) \cdot \delta x$,
where $x_0$ and $\delta_x = x - x_0$ are the atomic distance
in the equilibrium and the atomic displacement, respectively.
Such interaction of spin and lattice degrees of freedom
which conserves the spin quantum number
is described in a generic form
\begin{equation}
   {\cal H}_{\rm int} = \sum_{kq} \left\{
      g(k,q) a_{k+q}^\dagger a_k b_q  + h.c.
   \right\},
    \label{MPVertex1}
\end{equation}
where $a$ ($b$) is the magnon (phonon) annihilation operator
and $g$ is the magnon-phonon interaction vertex function.
The vertices are illustrated in Fig.~\ref{FigVertex}.
We define the magnon dispersion by $\omega_k$ which crosses the phonon
dispersion $\Omega_q$ at generic points of the brillouine zone 
denoted by $k^*$.

\begin{figure}[htb]
\hfil\epsfxsize=6cm\epsfbox{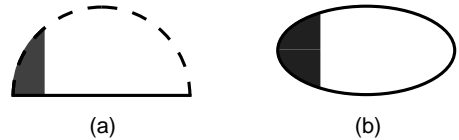}\hfil
\caption{Self-energy diagrams for (a) magnon
and (b) phonon. Shaded areas represent vertex corrections.}

\label{FigDiagram}
\end{figure}

{\em Magnon linewidth:}
Linewidth of magnons is calculated
from the self-energy diagram in Fig.\ref{FigDiagram}(a).
From the lowest order perturbation calculation
at $T=0$, magnon linewidth broadening $\Gamma_{\rm mag}$
due to the vertex (\ref{MPVertex1})
is given by
\begin{equation}
  \Gamma_{\rm mag}(k) = {\sl Im} \Pi_{\rm mag}(k,\omega_k)
  \propto \sum_q \delta( \omega_k - \omega_{k-q} - \Omega_q),
  \label{defGammaMag}
\end{equation}
where $\Pi_{\rm mag}$ is the magnon self-energy.

\begin{figure}[htb]
\hfil\epsfxsize=7cm\epsfbox{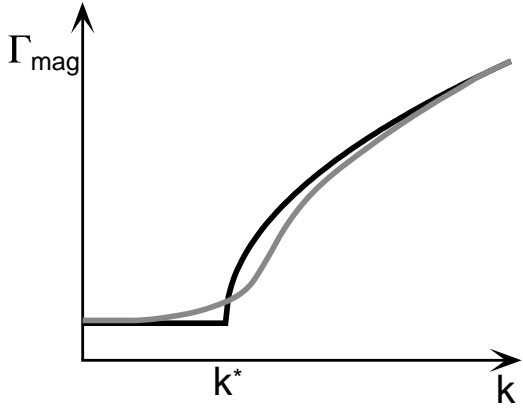}\hfil
\caption{Magnon linewidth as a function of wavevector $k$.
Magnon branch and phonon branch cross at $k^*$.
Solid curve is for the case of dispersionless Einstein phonon,
where three dimensional magnon density of states is assumed.
Grey curve schematically
represents the linewidth of magnons coupled to
optical phonons with finite dispersions.}

\label{FigWidth}
\end{figure}

If we consider an Einstein phonon $\Omega_q = \Omega_0$, we have
\begin{equation}
  \Gamma_{\rm mag}(k) \propto  D_{\rm mag} ( \omega_k - \Omega_0)
\end{equation}
where $D_{\rm mag}$ is the magnon density of states.
In three dimension, $D_{\rm mag}(\omega) \sim \sqrt{\omega}$ for $\omega\ge0$
while $D_{\rm mag}=0$ for $\omega<0$.
The result is schematically illustrated in Fig.~\ref{FigWidth}.
Namely, for $\omega_k < \Omega_0$  the linewidth of the magnon 
due to the magnon-phonon interaction is small.
As the magnon dispersion crosses the phonon branch,
the linewidth suddenly becomes broad.
In the case of finite dispersion of phonon frequency,
the onset of the linewidth broadening at 
the magnon-phonon crossing point
will be smeared out, as is indicated by the grey curve
in Fig.~\ref{FigWidth}.
For two-dimensional magnons, the density of states is
step-function like, and a more abrupt increase of magnon linewidth
is expected.

The result is intuitively understood as follows.
When magnon frequency is larger than the crossing frequency,
it is possible that a magnon with momentum and frequency ($k$, $\omega_k$) 
is  scattered into a magnon ($k-q$, $\omega_{k-q}$) and 
a phonon ($q$, $\Omega_q$)
obeying $\omega_k = \omega_{k-q} + \Omega_q$ through the magnon-phonon
interaction (\ref{MPVertex1}). Thus magnon energy dissipates
into the bosonic bath of optical phonons.
Near the zone center $k \ll k^*$, magnons with smaller frequencies
do not suffer from  such processes.

In the strong coupling region, 
a self-consistent treatment of phonons and magnons
gives a qualitatively similar result. Finite lifetime of magnons and phonons
creates the broadening of the delta-function in (\ref{defGammaMag})
which again causes a smearing of the onset of the magnon broadening.
At the onset, relevant contributions come from small frequency
magnons $\omega_k$ at $k\sim0$ and phonons at the crossing point $q\sim k^*$.
Therefore the broadening of the delta-function should not be large.
An incoherent part of the propagators creates a weakly $q$-dependent background
in magnon width.

{\em Phonon linewidth:}
Phonon linewidth is calculated
from the self-energy diagram in Fig.\ref{FigDiagram}(b).
Note that an magnon-phonon interaction which conserves
spin quantum number creates a self-energy with only
magnon-antimagnon diagrams.
In this case, we have
\begin{equation}
  \Gamma_{\rm ph}(q) \propto 
   \sum_k  \delta( \Omega_q - \omega_{k+q} + \omega_{k})
           \cdot ( n(\omega_{k}) - n(\omega_{k+q}) ) ,
	   \label{defGammaPh}
\end{equation}
where $n(\omega)$ is the Bose distribution function.
At $T\to 0$, we have $n(\omega_k)\to 0$ so that $\Gamma_{\rm ph}\to 0$.
Thus, the difference in the type of self-energy diagram
(Fig.~\ref{FigDiagram}(a-b)) explains
that phonon linewidth does not show broadening
even if the magnon-phonon interaction is strong enough
to exhibit anomalous magnon broadening.
Only at finite temperature there is a finite contribution
from the thermal spin fluctuation.

In the presence of spin-orbit ($L$-$S$) 
coupling, spin quantum number is
not conserved. In such a case, magnon-phonon hybridization terms
{\em e.g.}  $\sum \lambda(k) (a_k^\dagger b_k + h.c.)$
appear.\cite{Lovesey84}
These terms create hybridization gap at the magnon-phonon crossing point.
Since in high-$T_{\rm c}$ (La,Sr)MnO$_3$ 
a smooth crossing of magnon and phonon branch
is experimentally observed,\cite{Moudden98}
 such a hybridization term should be small,
although the anomalous Hall effect
in the ferromagnetic phase\cite{Asamitsu98,Matl98}
suggests it be non-zero.\cite{Kondo62}

\begin{figure}[htb]
\hfil\epsfxsize=7cm\epsfbox{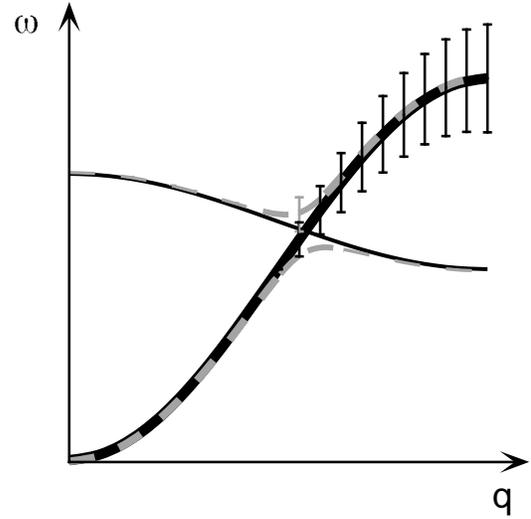}\hfil
\caption{Dispersion relation and damping in absence (solid curve)
and presence (grey dashed curve) of the magnon-phonon hybridization.
Thick (thin) curve is for a branch with a major
magnon (phonon) character. Error bars represent the magnon linewidth.}

\label{FigDispersion}
\end{figure}

Since the hybridization terms are bilinear in magnon and phonon operators,
this interaction does not cause the magnon broadening effect.
Hybridization causes the mixture of magnon and phonon
wavefunctions in both excitation spectra.
If hybridization is small, as expected, the
qualitative behavior in the linewidth does not change.
Broadening suddenly occurs at the ''optical'' branch
above the crossing point which is mostly magnon-like but has
non-zero phonon character as well.
In Fig.~\ref{FigDispersion} we schematically illustrate the 
dispersion relation of the magnon and phonon branches
as well as the magnon linewidth in the presence and absence
of magnon-phonon hybridization terms.

{\em Discussion:}
A crucial test for the present model is to measure the
temperature dependence of the phonon linewidth.
From eq.~(\ref{defGammaPh}) we expect an  increase of the
linewidth as temperature is increased, through the
magnon population $n(\omega_k)$.
At a fixed temperature, $n(\omega_k)$ is maximum at $k\to0$ ($\omega_k\to0$)
and eq.~(\ref{defGammaPh}) suggests that $\Gamma_{\rm ph}$
is maximum at $\Omega_q \simeq \omega_{\rm q}$.
Namely, the phonon linewidth should be maximum around the crossing point.
At the Curie temperature $T_{\rm c}$, magnon population should
critically increase, so yet another
 anomalous broadening is expected in the phonon linewidth.

The magnon damping effect given here can be generalized
to any system where magnons couple with other bosonic
degrees of freedom through the interaction in the form (\ref{MPVertex1}).
In manganites, another possible source of damping is the
orbital wave excitations.\cite{Ishihara96}
Namely, accurate measurement of magnon dispersion and linewidth
might become an indirect method to investigate the orbital dynamics.
At the same time, for  metallic manganites in the
 low-temperature ferromagnetic phase
at around $x\sim 0.3$,
 no anomalous magnon broadening has been reported so far in the
low energy region $\simle 10{\rm meV}$. This 
might indicate that it is not likely to assume  
low-energy orbital fluctuations.\cite{Ishihara97}
The mechanism for magnon broadening discussed here 
does not require an intrinsic modification to the picture of
half-metallic double exchange ferromagnet as the
low temperature state of manganites in the low energy region.

By decreasing A-site radius in manganites,
increase of the magnon damping at the zone boundary
is prominent, and at the same time
spin stiffness constant which is a measure for
electron kinetic energy\cite{Kubo72,Furukawa96} only makes a small
change.\cite{Dai99x}
Similarly, decrease of $T_{\rm c}$ by A-site ionic radius control
has been known to be unexpectedly large compared to the
nominal bandwidth change.\cite{Hwang95}
Therefore, it is not likely to simply assume an
increase of magnon-phonon coupling through
the decrease of electronic bandwidth while keeping the lattice couplings
constant. A possible origin of a sudden increase in the coupling constant
might be due to the fact that the metallic state is in a close vicinity
to the charge ordered insulating state, through 
fluctuations toward the phase  with larger  lattice distortions.
Note that such increase is speculated to occur both at low temperature and
at $T\sim T_{\rm c}$, although
it has been considered that
dynamical lattice distortions are suppressed
in the low temperature ferromagnetic phase.\cite{Millis96l}
Further study of such dynamical fluctuations may give more
clear understandings in the intermediate compounds at low temperature,
including a possible phase separation phenomena toward
ferromagnetic metal and charge ordered insulator.\cite{Yunoki98}

To summarize, the effects of the magnon-phonon interaction 
to the broadening of magnons are studied
at the presence of magnon-phonon dispersion crossing.
When the interaction conserves spin quantum number,
the magnon linewidth becomes broad only when magnon frequency
is larger than the magnon-phonon crossing frequency.
Anomalous broadening at the zone boundary,
which is observed in the narrow bandwidth CMR manganites,
will be reproduced if we simply assume that
the magnon-phonon coupling is strong enough.
Phonon linewidth does not show broadening at low temperatures.
A crucial test for this model is proposed.
The author thanks J. Fernandez-Baca and P. Dai for discussion.
This work was supported by Mombusho Grant-in-Aid No.~11640365.

%\bibliographystyle{jpsj}
%\bibliography{all,local}

\end{document}